\documentclass{svjour2}
\usepackage{graphicx}

\journalname{General Relativity and Gravitation}

\def\citen#1{\cite{#1}}
\def\beq{\begin{equation}}
\def\eeq{\end{equation}}
\def\rmd{{\rm d}}

\def\vecA{\vec a(n)} \def\A{a(n)}

\begin{document}

\title{Physical frames along circular orbits in stationary axisymmetric spacetimes}

\author{Donato Bini \and
        Christian Cherubini \and
        Andrea Geralico  \and
        Robert T. Jantzen
}

\institute{Donato Bini 
              \at
              Istituto per le Applicazioni del Calcolo ``M. Picone,'' CNR I-00161 Rome, Italy\\
              ICRA, University of Rome ``La Sapienza,'' I--00185 Rome, Italy\\
              INFN - Sezione di Firenze, Polo Scientifico, Via Sansone 1, I--50019, Sesto Fiorentino (FI), Italy\\
              \email{binid@icra.it} 
    \and
              Christian Cherubini 
              \at
              Facolt\'a di Ingegneria, Universit\`a Campus Biomedico, Via E. Longoni 83,  I-00155 Roma, Italy\\
              ICRA, University of Rome ``La Sapienza,'' I--00185 Rome, Italy\\
              \email{cherubini@icra.it}         
    \and
          Andrea Geralico 
              \at
              Physics Department and ICRA, University of Rome ``La Sapienza,'' I--00185 Rome, Italy\\
              \email{geralico@icra.it}
    \and
          Robert T. Jantzen 
              \at
              Department of Mathematical Sciences, Villanova University, Villanova, PA 19085,  USA\\
              ICRA, University of Rome ``La Sapienza,'' I--00185 Rome, Italy\\
              \email{robert.jantzen@villanova.edu}
}

\date{Received: date / Accepted: date / Version: date }

\maketitle

{\it Dedicated to Bahram Mashhoon for his 60th birthday}

\begin{abstract}
Three natural classes of orthonormal frames, namely Frenet-Serret, Fermi-Walker and parallel transported frames,
exist along any timelike world line in spacetime. Their relationships are investigated for timelike circular orbits in stationary axisymmetric spacetimes, and illustrated for black hole spacetimes.
\end{abstract}

\section{Introduction}

As the simplest generalizations of periodic Newtonian orbits,
timelike circular orbits in stationary axisymmetric spacetimes have been been the subject of intense interest over the years for various reasons. They have been used
i) to investigate relativistic generalizations of inertial forces in relativity and the special properties of circular orbits in this context 
\cite{acl1988,al1997,mfg,fdfacc,semerak,idcf1,idcf2,page} 
(see the summary in Ref.~\citen{inertialforces}), 
ii) to study the effects of frame dragging  (locally nonrotating and static observers, gyroscope precession, clock-effects) \cite{mtw,rind-perl,tart,bj,bjm}, 
iii) to investigate the holonomy invariance of vectors undergoing parallel transport along them \cite{roth,maartens,bjm_chce,bcj,bccj,bjnc}, and
iv) and to discuss the geometry of embedding diagrams associated with them which provide a visualizable interpretation for the spatial geometry contribution to curvature effects (see Ref.~\citen{bjed} for references).
The mathematical tools used for the description of the properties of such highly symmetric orbits include standard tensorial techniques, spacetime splitting techniques (``gravitoelectromagnetism"), the Newman-Penrose formalism, and the Frenet-Serret intrinsic framework.

Among the family of all possible such individual timelike circular orbits, some members are privileged by special properties.
There exist certain special timelike orbits satisfying ``intrinsic properties" like the vanishing of the Frenet-Serret curvature or torsions (like the geodesics or extremely accelerated observers, for example) or  ``extrinsic properties"  due to the background geometry in which they are considered, notably properties relative to the symmetry of the spacetime (most familiar in black hole spacetimes, where the static observers and the zero angular momentum observers each play key roles, for example).
A congruence of such similarly special orbits in a given spacetime can be thought of as the world lines of a family of preferred observers which can be used to interpret the spacetime geometry by measuring tensors in terms of their projected spatial/temporal parts.  It is then natural to introduce adapted spatial orthonormal triads suited for such measurements, taking into account that a triad which may seem convenient from a geometrical point of view may also be absolutely nontrivial to construct in practice. The most important characteristic of such a spatial triad is how it is transported along the orbit. These triads are usually either Fermi-Walker transported (which reduces to parallel transport for geodesics) or are Frenet-Serret frames whose orientation is determined by the intrinsic differential properties of the orbits themselves. Calling such frames ``physical" seems appropriate since they can be interpreted directly in terms of local physical quantities, like the acceleration direction or the directions of test gyroscopes along a given world line. Parallel transported orthonormal frames along accelerated orbits differ from both Fermi-Walker and Frenet-Serret frames, and do not remain adapted to the local rest space of the orbits, but are also of interest since
they determine the curvature properties of spacetime along such world lines. However, the additional boost they undergo relative to the local rest space of the orbit itself makes them more complicated to describe.

Here we evaluate explicitly these three types of spatial triads for timelike circular orbits in stationary axisymmetric spacetimes in general and for the important special case of the Schwarzschild and Kerr black hole spacetimes.
It is remarkable that while Frenet-Serret frames have been studied in a great detail in this context \cite{iyer-vish,circfs},
explicit expressions for Fermi-Walker frames and parallel transported frames (apart from the case of geodesic orbits \cite{marck1,marck2} in which they coincide) are difficult to find in the literature even for Minkowski spacetime \cite{caipap,bj-rrf} where they may not even be identified as such but merely chosen as ``convenient" (see also Ref.~\citen{bjnc}). It is exactly the nongeodesic orbits which underly the symmetry adapted coordinates in which the spacetime geometry for these spacetimes is most simply expressed, and therefore while they are not interesting as freely falling particle paths, they do offer insights into how the spacetime geometry manifests itself. Nongeodesic circular orbits like points fixed on the rotating earth also play an important role in GPS, as do Fermi coordinate systems built on Fermi-Walker transported frames along them (see Ref.~\citen{bahder} for an exhaustive current list of references for Fermi coordinates in the context of circular orbits).

The mathematical machinery needed to obtain explicit representations of these frames along circular orbits in this symmetry class of spacetimes involves the Lie algebra of the Lorentz group and its transformation properties under a change of basis that mirrors the transformation properties of the electric and magnetic field vectors associated with an electromagnetic 2-form, already extensively described elsewhere in the study of parallel transport around circular orbits \cite{bccj}. This in turn is related to the eigenvector method of solving a system of first order linear homogeneous constant coefficient differential equations, needed to go from the Lie algebra of the Lorentz group to the finite group elements via the exponential map and actually carry out the matrix exponentiation. The present article completes and extends the work of \cite{bccj} on parallel transport to yield explicit expressions for not only the parallel transported frames but also the Fermi-Walker frames. 

\section{Circular orbits and adapted frames in stationary axisymmetric spacetimes}

A brief review of the description of this scenario follows.

\subsubsection*{Stationary axisymmetric spacetimes} 

Using a coordinate system $\{t,r,\theta,\phi \}$ adapted to the spacetime symmetries, i.e., with $\partial_t$ (timelike) and $\partial_\phi$ (spacelike, with closed coordinate lines) a pair of commuting Killing vectors, the metric can be expressed by a line element of the form
\beq
\label{metr_gen}
\rmd s^2= g_{tt} \rmd t^2 + 2 g_{t\phi} \rmd t \rmd \phi + g_{\phi\phi}\rmd \phi^2 + g_{rr}\rmd r^2 + g_{\theta\theta} \rmd \theta^2\ ,
\eeq
where all the metric coefficients depend only on $r$ and $\theta$, provided that the metric belongs to the most interesting class of orthogonally symmetric such metrics \cite{ES}. The time coordinate lines, when timelike, are the world lines of the static observers.

\subsubsection*{ZAMOs}

The zero angular momentum observer (ZAMO) family of fiducial observers has a 4-velocity $n$ characterized as that normalized linear combination of the two given Killing vectors which is orthogonal to $\partial_\phi$ and future-pointing, and it is the unit normal to the time coordinate hypersurfaces
\beq
\label{n}
n=N^{-1}(\partial_t-N^{\phi}\partial_\phi)\ ,
\eeq
where $N=(-g^{tt})^{-1/2}$ and $N^{\phi}=g_{t\phi}/g_{\phi\phi}$ are the lapse function and only nonvanishing component of the shift vector field respectively. 
Our discussion is limited to those regions of spacetime where the time coordinate hypersurfaces are spacelike: $g^{tt}<0$. 
A suitable orthonormal frame adapted to the ZAMOs and invariant under the symmetry group action is given by
\beq
\label{zamoframe}
e_{\hat t}=n\ , \quad
e_{\hat r}=\frac1{\sqrt{g_{rr}}}\partial_r\ , \quad
e_{\hat \theta}=\frac1{\sqrt{g_{\theta \theta }}}\partial_\theta\ , \quad
e_{\hat \phi}=\frac1{\sqrt{g_{\phi \phi }}}\partial_\phi\ ,
\eeq
with dual 
\beq
\omega^{{\hat t}}=N\rmd t\ , \quad \omega^{{\hat r}}=\sqrt{g_{rr}}\rmd r\ , \quad 
\omega^{{\hat \theta}}= \sqrt{g_{\theta \theta }} \rmd \theta\ , \quad
\omega^{{\hat \phi}}=\sqrt{g_{\phi \phi }}(\rmd \phi+N^{\phi}\rmd t)\ ,
\eeq
so that the line element (\ref{metr_gen}) can be also expressed in the form
\beq
\rmd s^2 = -N^2\rmd t^2 +g_{\phi \phi }(\rmd \phi+N^{\phi}\rmd t)^2 + g_{rr}\rmd r^2 +g_{\theta \theta}\rmd \theta^2\ . 
\eeq

The accelerated ZAMOs are locally nonrotating in the sense that their vorticity vector $\omega(n)$ vanishes, but they have a nonzero expansion tensor $\theta(n)$ whose nonzero
components can be completely described by the shear vector
$\theta_{\hat \phi}(n)^\alpha = \theta(n)^\alpha{}_\beta\,{e_{\hat\phi}}^\beta$. The contravariant components (notationally ignoring the distinction between index-shifted tensors if indices are not present) are
\beq
\theta(n) = e_{\hat\phi}\otimes\theta_{\hat\phi}(n)
           +\theta_{\hat\phi}(n)\otimes e_{\hat\phi}\ .
\eeq 
Since the expansion scalar (its trace) is zero, the expansion and shear tensors coincide.

The nonzero ZAMO kinematical quantities (acceleration $a(n)=\nabla_n n$ and shear tensor) and the conveniently defined Lie relative curvature vector \cite{idcf2,bjdf} only have nonzero components in the $r$-$\theta$ 2-plane of the tangent space, convenient to call the acceleration plane
\begin{eqnarray}
\label{accexp}
\vecA & = & [a(n)^{\hat r},\,a(n)^{\hat\theta}]
 =[\partial(\ln N)/\partial\hat r ,\,\partial(\ln N)/\partial\hat\theta]
=\vec\nabla(n)\ln N\ ,
\nonumber\\
\vec\theta_{\hat\phi}(n) 
& = &[\theta_{\hat\phi}(n)^{\hat r},\, \theta_{\hat\phi}(n)^{\hat\theta}]
  = -\frac{\sqrt{g_{\phi\phi}}}{2N}\,[\partial N^\phi/\partial\hat r,\,\partial N^\phi/\partial\hat\theta]
\nonumber\\
&=& -\frac{\sqrt{g_{\phi\phi}}}{2N} \vec\nabla(n)N^\phi\ , 
\nonumber\\
\vec k_{(\rm lie)}(n)
& = & [k_{(\rm lie)}(n)_{\hat r},\,k_{(\rm lie)}(n)_{\hat\theta}]
 = -[\partial(\ln \sqrt{g_{\phi\phi}})/\partial\hat r ,\,\partial(\ln \sqrt{g_{\phi\phi}})/\partial\hat\theta]
\nonumber\\
&=&-\vec\nabla(n)\ln(\sqrt{g_{\phi\phi}})\ ,
\end{eqnarray}
where a boldface 2-vector notation for pairs of orthonormal frame components of vectors belonging to this subspace has been conveniently adopted.
Here $\partial_{\hat r}\equiv e_{\hat r}$ and $\partial_{\hat \theta}\equiv e_{\hat \theta}$, while the spatial covariant derivative $\nabla(n)=P(n)\nabla$ is obtained by projecting $\nabla$ onto the local rest space of $n$ using the spatial projector $P(n)^\alpha{}_\beta=\delta^\alpha{}_\beta+n^\alpha n_\beta$ \cite{mfg}.
In the static limit $N^\phi\to0$, the shear vector $\vec\theta_{\hat\phi}(n)$ vanishes.

\subsubsection*{Uniformly rotating world lines}

The family of uniformly rotating timelike circular orbits at a given fixed location $(r,\theta)$ consists of Killing trajectory helices confined to the $(t,\phi)$ coordinate cylinders (symmetry group orbits). These world lines have
a 4-velocity vector $U$ which belongs to the Killing 2-plane tangent to this cylinder. It can be parametrized equivalently either by the constant angular velocity $\zeta$ with respect to infinity (just the slope $\rmd\phi/\rmd t$ on this coordinate cylinder)
or by the constant relative velocity $\nu$ with respect to the ZAMOs (defining the usual gamma factor $\gamma=(1-\nu^2)^{-1/2} $) or by the constant ZAMO rapidity $\alpha$ as follows 
\beq
\label{orbita}
U=\Gamma [\partial_t +\zeta \partial_\phi ]
 =\gamma [n +\nu e_{\hat \phi}]
 =\cosh\alpha \, n + \sinh\alpha \, e_{\hat\phi}
\ ,
\eeq
where $\Gamma$ is a normalization factor such that $U_\alpha U^\alpha =-1$ and hence
\beq
\Gamma =\left[ N^2-g_{\phi\phi}(\zeta+N^{\phi})^2 \right]^{-1/2}
       =\gamma/N
\eeq
with
\beq
\zeta=-N^{\phi}+(N/\sqrt{g_{\phi\phi}})\, \nu\ ,\quad
\nu=\sqrt{g_{\phi\phi}} (\zeta+N^{\phi})/N =\tanh \alpha\ .
\eeq
It is useful to introduce a spacelike unit vector $\bar U$ within the Killing 2-plane which is orthogonal to $U$ given by
\beq
\label{barU}
\bar U=\bar \Gamma [\partial_t +\bar \zeta \partial_\phi ]\ ,
\eeq
with
\beq
\bar \zeta=-\frac{g_{tt}+\zeta g_{t\phi}}{g_{t\phi}+\zeta g_{\phi\phi}}
  =-N^{\phi}+(N/\sqrt{g_{\phi\phi}})\, \nu^{-1}\ , \qquad 
\bar \Gamma=\Gamma\nu\ .
\eeq
This vector is aligned with the azimuthal direction in the local rest space of $U$, with direction depending on the sign of $\bar\zeta$, and is therefore a spatial normal to the acceleration plane, which contains the acceleration vector of the circular orbit
\beq
a(U)= \gamma^2 [ a(n) +2\nu\, \theta_{\hat \phi}(n) +\nu^2 k_{\rm(lie)}(n) ]\ .
\eeq

Note that the azimuthal coordinate $\phi$ along the orbit depends on the coordinate time $t$ or proper time $\tau$ along that orbit according to 
\beq\label{eq:phitau}
  \phi-\phi_0  = \zeta (t-t_0) = \Omega_U (\tau_U-\tau_{U0}) \ ,\quad
\Omega_U =\Gamma\zeta
\ ,
\eeq
defining the corresponding coordinate and proper time orbital angular velocities $\zeta$ and $\Omega_U$. These relations determine the rotation of the spherical frame with respect to a nonrotating frame at infinity. Notice that introducing the static family of observers at rest with respect to the coordinates which have $\zeta=0$,  
their relative velocity with respect to ZAMOs is
\beq
\nu_0=\sqrt{g_{\phi\phi}}\, N^\phi/N
\eeq
and $\Omega_U$ can also be written as
\beq
\Omega_U=(\gamma/N)(\nu-\nu_0)\ .
\eeq

Note that when one considers axes which are nonrotating with respect to infinity, one must undo the rotation of the spherical axes which rotate with respect to infinity with the orbital angular velocity when it is nonzero. For example, in the most interesting case of the equatorial plane $\theta=\pi/2$, one can introduce explicitly the local axes  $e_x$, $e_y$ and $e_z$ which are seen as nonrotating with respect to infinity by rotating the spherical axes in the plane of the orbital rotation by an angle 
$\phi_\infty=\Omega_U (\tau_U-\tau_{U0})=\phi-\phi_0$ 
\beq\label{nonrotframeinfinity}
  e_x = \cos(\phi_\infty) e_{\hat r} -\sin(\phi_\infty) e_{\hat\phi}\ ,\
  e_y = \sin(\phi_\infty) e_{\hat r} +\cos(\phi_\infty) e_{\hat\phi}\ ,\
  e_z = -e_{\hat\theta}\ ,
\eeq
which is a clockwise rotation when $\Omega_U>0$ and counterclockwise when $\Omega_U<0$.

\section{Orthonormal frames and transport}

The symmetry adapted frame has constant components for the induced connection matrix along any given circular orbit
\beq
\frac{D e_{\hat\alpha}}{d\tau_U} = e_{\hat\beta} \Gamma^{\hat\beta}{}_{\hat\gamma\hat\alpha} U^{\hat\gamma} 
= e_{\hat\beta} F^{\hat\beta}{}_{\hat\alpha}
\eeq
since the connection components depend only on $r$ and $\theta$ while the frame components of the 4-velocity are constants.
This induced connection matrix $F^{\hat\beta}{}_{\hat\alpha}$, which is antisymmetric (when index-lowered) and constant along the curve, generates a Lorentz transformation along the curve whose inverse action on the symmetry adapted orthonormal frame can be used to transform it to a parallel transported orthonormal frame.

Among all orthonormal frames defined along the curve, those which are transformable from the symmetry adapted frame by a constant Lorentz transformation
\beq
E_\alpha = e_{\hat\beta} L^{\hat\beta}{}_\alpha\ ,\qquad
d L^{\hat\beta}{}_\alpha/d\tau_U=0
\eeq
are such that the induced connection transforms as an ordinary tensor. Thus one may reduce the induced connection matrix to special forms in analogy with transforming a single electromagnetic field 2-form at a point, a discussion which then depends on the invariants of this matrix under such transformations \cite{bcj}. 
In the general case the Lorentz transformation along the orbit which determines parallel transport 
represents simultaneous boosts and rotations in an orthogonal pair of 2-planes, and one must first orient a frame to this orthogonal decomposition to obtain explicit formulas for a frame which then undergoes these boosts and rotations.

In particular the Frenet-Serret frame $\{E_\alpha\}$ defined along  a circular orbit is related to the symmetry adapted orthonormal frame by a constant Lorentz transformation, and may be constructed by a simple process of repeated differentiation and orthogonalization, with the Frenet-Serret curvatures and torsions parametrizing the induced connection matrix.
The result turns out to be a constant Lorentz transformation of the symmetry adapted frame.
Starting from this intrinsically defined frame, one can then remove the spatial rotation relative to a Fermi-Walker transported frame by a time-dependent rotation defined along the curve associated with the constant Frenet-Serret angular velocity to obtain a Fermi-Walker transported frame explicitly. To obtain a parallel transported frame starting from the Frenet-Serret frame, one must instead simultaneously remove a time-dependent boost and a time-dependent rotation, each with constant rates, after first re-orienting the axes with a constant boost and rotation which adapts the frame to the pair of orthogonal 2-planes in which these boosts and rotations occur. Of course this last frame is no longer adapted to the local space and time decomposition of $U$ if the curve is not a geodesic. The underlying mathematical setting for these calculations is just that of constant coefficient linear systems of first order differential equations, where the coefficients are symmetry invariant functions determined by the spacetime geometry. They really just involve the geometry of the exponential map from the Lie algebra of the Lorentz group to the group manifold.
By first adapting the symmetry invariant frame to the real eigenvectors and real and complex parts of the complex eigenvectors of the Lie algebra-valued coefficient matrix, one can easily express the one-parameter group of Lorentz transformations which solves the problem.

\subsection{Frenet-Serret transport}

The spacetime Frenet-Serret frame $\{E_\alpha\}$ $(\alpha=0,1,2,3)$ adapted to the circularly rotating orbit with unit tangent vector $U=E_0$ is described by the following system of transport equations \cite{iyer-vish}
\begin{eqnarray}
\label{FSeqs}
\frac{DE_0}{d\tau_U}&=\kappa E_1\ ,\phantom{+\tau_1 E_2\ \ \ } \qquad &  
\frac{DE_1}{d\tau_U}=\kappa E_0+\tau_1 E_2\ ,\nonumber \\
 \nonumber \\
\frac{DE_2}{d\tau_U}&=-\tau_1E_1+\tau_2E_3\ , \qquad &
\frac{DE_3}{d\tau_U}=-\tau_2E_2\ .
\end{eqnarray} 
This is conveniently described in terms of the constant matrix of the induced connection along the world line
\beq\label{eq:F}
\frac{D E_\alpha}{d\tau_U} = E_\beta F^\beta{}_\alpha\ , \qquad
(F^\alpha{}_\beta)
= \left(\begin{array}{cccc}0&\kappa&0&0\\ \kappa& 0&-\tau_1&0\\ 0&\tau_1&0&-\tau_2\\ 0&0&\tau_2&0\end{array}\right)\ .
\eeq
Apart from the optional sign, the curvature $\kappa$ is the magnitude $||a(U)||$ of
the acceleration $a(U)\equiv DU/d\tau_U=\kappa E_1$, while
the first
and second torsions $\tau_1$ and $\tau_2$ are the components of the Frenet-Serret angular velocity vector
\beq
\label{omegaFS}
\omega_{\rm (FS)}=\tau_1 E_3 + \tau_2 E_1\ , \qquad ||\omega_{\rm (FS)}||=[\tau_1^2 + \tau_2^2]^{1/2}\ ,
\eeq
putting the spatial transport equations in the form
\beq\label{eq:DXdtau}
\frac{DE_a}{d\tau_U}=\omega_{\rm (FS)}\times_U E_a+\kappa E_0\,\delta^1_a\ .
\eeq
The curvature boosts the orthonormal frame relative to parallel transport so that the frame remains aligned with $U$, while the angular velocity rotates the spatial frame relative to parallel transport.

In the case of circular orbits and stationary axisymmetric spacetimes all Frenet-Serret curvature and torsions are constant along the orbit and 
the problem of defining such a frame has been explicitly solved by Iyer and Vishveshwara in 1993 \cite{iyer-vish}.
Their results can be reformulated  \cite{bjdf} as follows. The Frenet-Serret vectors are related to the ZAMO observer adapted frame by a boost in the Killing 2-plane and a rotation in the acceleration plane
\begin{eqnarray}
\label{FSframe}
E_0 &=&\cosh \alpha\, n+\sinh \alpha\, e_{\hat \phi} \ ,\qquad
E_1=\cos \chi\, e_{\hat r}+\sin \chi\, e_{\hat \theta}\ ,\ 
\nonumber \\
E_2&=&\sinh \alpha\, n +\cosh \alpha\, e_{\hat \phi}\ ,\qquad 
E_3=\sin \chi\, e_{\hat r}-\cos\chi\, e_{\hat \theta}\ ,
\end{eqnarray}
where  $E_2=\rmd U/\rmd\alpha\equiv \bar U$, $E_3=-\rmd e_1/\rmd\chi$ and polar coordinates $(\kappa, \chi)$ have been introduced in the acceleration plane according to 
\begin{equation}
a(U)=a(U)^{\hat r}e_{\hat r}+a(U)^{\hat \theta}e_{\hat \theta}
=\kappa E_1\ , \
a(U)^{\hat r}=\kappa \cos\chi\ , \ 
a(U)^{\hat \theta}=\kappa \sin\chi\ .
\end{equation} 
The expressions for the two torsions are given by
\beq\label{eq:torsion}
\tau_1=-\frac12 \frac{\rmd\kappa}{\rmd\alpha}
=-\frac{1}{2\gamma^2} \frac{\rmd\kappa}{\rmd\nu}\ , \qquad 
\tau_2=-\frac12 \kappa\frac{\rmd \chi}{\rmd\alpha}
=-\frac{\kappa }{2\gamma^2} \frac{\rmd\chi}{\rmd\nu}\ .
\eeq
A Frenet-Serret frame is not in general Fermi-Walker or parallel propagated along $U$. However,
starting from a Frenet-Serret frame, one can then construct these other frames along the orbit.

The spatial Frenet-Serret frame is phase-locked to the orbit itself, according to a terminology first introduced by de Felice 
\cite{fdfacc} to indicate that apart from the boost between the local rest frames of $n$ and $U$, on each orbit the spatial Frenet-Serret frame is related to the symmetry adapted spatial frame 
$\{e_{\hat r}, e_{\hat\theta}, e_{\hat\phi} \}$ by a fixed rotation along the orbit, since this angle $\chi$ depends only on $r$ and $\theta$.

\subsection{Fermi-Walker transport} 

A vector $X$ undergoes Fermi-Walker transport along $U$ if its Fermi-Walker derivative vanishes
\beq
\label{fw}
\nabla_{({\rm fw}, U)} X^\alpha
\equiv \frac{DX^\alpha}{d\tau_U}+[a(U)\wedge U]^\alpha{}_\beta X^\beta =0\ .
\eeq
Fermi-Walker transport is just the boost in the acceleration plane along the trajectories of $U$ needed to keep $U$ aligned with itself, reflected in the relation $\nabla_{({\rm fw}, U)} U=0$, while
if $X$ is orthogonal to $U$, i.e., $X\cdot U=0$, then this simplifies to
\beq
\nabla_{({\rm fw}, U)} X^\alpha 
= \frac{DX^\alpha}{d\tau_U}-\kappa X_\beta E^\beta_1\, E^\alpha_0=0\ ,
\eeq
and can be solved for a triad of orthonormal spatial vectors to join $U$ in forming an orthonormal frame undergoing Fermi-Walker transport along $U$.
When explicitly expressed in terms of the Frenet-Serret frame, this becomes
\beq
  \frac{d X^\alpha}{d\tau_U} + (F|_{\kappa=0})^\alpha{}_\beta X^\beta=0\ ,
\eeq
which corresponds to a pure rotation 
\beq
   X^\alpha(\tau_U) = [e^{-\tau_U (F|_{\kappa=0})}]^\alpha{}_\beta X^\beta(0)
\eeq
in the local rest space of $U$ about the
Frenet-Serret angular velocity vector $\omega_{\rm (FS)}$ which is itself Fermi-Walker transported along $U$ since the torsions are constant along $U$ (so the derivative of the linear combination is the linear combination of the derivatives) and by Eq.~(\ref{eq:DXdtau}). Thus this angular velocity is only boosted and not rotated relative to a parallel transported frame.

By first aligning the frame with $\omega_{\rm (FS)}$ by a constant counterclockwise rotation of the Frenet-Serret frame in the $E_1$-$E_3$ plane through an angle whose tangent is the ratio $-\tau_2/\tau_1$ 
and then rotating the resulting frame about that nonrotating axis with an angular velocity of opposite sign, one obtains a Fermi-Walker transported frame.
The first step leads to the new frame
\begin{eqnarray}
H_1 &=&(\tau_1 E_1-\tau_2 E_3)/||\omega_{\rm (FS)}|| \ ,\nonumber \\
H_2 &=&E_2 \ ,\nonumber \\
H_3  &=&\omega_{\rm (FS)}/||\omega_{\rm (FS)}||
      =(\tau_1 E_3+\tau_2 E_1)/||\omega_{\rm (FS)}|| \ .
\end{eqnarray}
A rotation of oppositely signed angular velocity about $H_3$, letting
$\Delta\tau_U=\tau_U-\tau_{U0}$, then yields
\begin{eqnarray}
\label{fwframe}
F_1 &=&\cos(||\omega_{\rm (FS)}||\Delta\tau_U) H_1 
     - \sin(||\omega_{\rm (FS)}||\Delta\tau_U) H_2 
\ ,\nonumber \\
F_2 &=&\sin(||\omega_{\rm (FS)}||\Delta\tau_U) H_1 
     + \cos(||\omega_{\rm (FS)}||\Delta\tau_U)H_2 
\ ,\nonumber \\
F_3  &=&H_3\ ,
\end{eqnarray}
which form a Fermi-Walker transported spatial frame along $U$. Note that one is still free to pick the zero of the proper time $\tau_U$ along a given world line, allowing an additional fixed rotation about the third frame vector, but then any fixed rotation of the final frame will not change its character as a Fermi-Walker transported frame adapted to the space-plus-time decomposition of the tangent space associated with the 4-velocity $U$.
A similar statement holds for the analogous situation for parallel transport below, except that one must give up adapting the frame to the 4-velocity, and then any fixed Lorentz transformation of the final frame is allowed.

\subsection{Parallel transport} 

A vector $X$ is parallel transported along $U$ if its covariant derivative along $U$ vanishes:
$D X/d\tau_U =0$, which in the Frenet-Serret frame becomes
\beq
  \frac{d X^\alpha}{d\tau_U} + F^\alpha{}_\beta X^\beta=0\ ,
\eeq
which corresponds to a combined boost plus rotation 
\beq
   X^\alpha(\tau_U) = [e^{-\tau_U F}]^\alpha{}_\beta X^\beta(0)\ .
\eeq 
In addition to removing the angular velocity of the Frenet-Serret frame with respect to a Fermi-Walker transported frame, one must now also simultaneously undo the boost of $U$ compared to parallel transporting any particular value of $U$ along its world line.

When $\tau_1=0$ (the case of the extremely accelerated world lines  \cite{fdfacc,semerak,idcf2}, see Eq.~(\ref{eq:torsion})), the acceleration and Frenet-Serret angular velocity vectors are both aligned along $E_1$ and the Frenet-Serret induced connection matrix ($F^\alpha{}_\beta)$ (see Eq.~(\ref{eq:F})) is in its canonical form representing the generator of a boost in the $E_0$-$E_1$ plane and a simultaneous rotation in the orthogonal $E_2$-$E_3$ plane which results from the matrix exponential $e^{\tau_U F}$. Applying the inverse of this time-dependent Lorentz transformation  to the Frenet-Serret frame itself yields a Fermi-Walker transported frame.
In this case the Frenet-Serret frame is just the real orthonormal frame naturally associated with a complex null frame \cite{ES} consisting of eigenvectors of the matrix ($F^\alpha{}_\beta)$ (related to each other by sums and differences divided by a real or purely imaginary scalar).

When instead $\tau_1\neq0$ one has to first perform a constant Lorentz transformation to a new frame which makes that entry zero as well for the new induced connection matrix, and then the former situation holds. This is accomplished by aligning the electric and magnetic parts of the Frenet-Serret matrix with a boost, which is only possible if also $\tau_2\neq0$, and then rotating the resulting frame to align  it with that new common direction.
First in analogy with an electromagnetic field under constant linear transformations at a point where starting from a nonzero pair of electric and magnetic fields, one can transform them to a new frame in which they are parallel, a simple constant boost maps the electric vector $a(U)=\kappa E_1$ and magnetic vector
 $\omega_{\rm (FS)}=\tau_1 E_3 + \tau_2 E_1$ parts of the induced connection matrix $F$ onto  a new common direction.
This is accomplished by performing a constant boost in the plane of $U=E_0$ and $E_2$ (i.e., a boost in the azimuthal direction)
\begin{eqnarray}
f_0= \cosh \tilde \alpha\, E_0 + \sinh \tilde \alpha\, E_2\ , \qquad 
f_2= \sinh \tilde \alpha\, E_0 + \cosh \tilde \alpha\, E_2\ , 
\end{eqnarray}
which mixes up the electric and magnetic vectors in the $E_1$-$E_3$ plane under the transformation, 
where the boost parameter is determined by the condition that they be proportional
\beq
\label{duealpha}
\tanh 2 \tilde \alpha = \frac{2\kappa \tau_1}{\kappa^2 + \tau_1^2+\tau_2^2}\equiv T \ ,
\eeq
and the following relations hold
$$
\tanh\tilde \alpha =\frac{1-\sqrt{1-T^2}}{T}\equiv \mathcal{T}\ , \quad 
\sinh\tilde \alpha =\frac{\mathcal{T}}{\sqrt{1-\mathcal{T}^2}}\ , \quad 
\cosh\tilde \alpha =\frac{1}{\sqrt{1-\mathcal{T}^2}}\ .
$$
Note that if one assumes $\kappa\geq0$, then the sign of $\tilde \alpha$ is the same as the sign of $\tau_1$, and that
when $\kappa=0$ (geodesic) or $\tau_1=0$ (extremely accelerated), this boost reduces to the identity so both must be nonzero for this step to be nontrivial.
 
Then continuing to assume that the second torsion $\tau_2\neq0$,
one must next perform a spatial rotation in the $E_1$-$E_3$ plane to align the common direction of the new electric and magnetic parts of the induced connection matrix with the new frame vector $f_1$
\begin{eqnarray}
f_1= \cos \Theta\, E_1 + \sin \Theta\, E_3\ , \qquad 
f_3= -\sin \Theta\, E_1 + \cos \Theta\, E_3\ , 
\end{eqnarray}
with
\beq
\tan\Theta
=\frac{\tau_2}{\kappa \coth \tilde \alpha-\tau_1}
=-\frac{\kappa \tanh \tilde \alpha-\tau_1}{\tau_2}
\ .
\eeq

Given this constant re-orientation of the time direction and the spatial axes adapted to the pair of orthogonal 2-planes in which the simultaneous boost and rotation take place under parallel transport along the circular orbit, one can now undo them together. Perform
a time-dependent boost in the plane of  $f_0$ and $f_1$ at a constant boost rate $\sigma_B$ which undoes the effects of the acceleration of $U$ and a simultaneous time-dependent rotation in the plane of $f_2$ and  $f_3$ with constant angular velocity $\sigma_R$
\begin{eqnarray}\label{parframe}
\epsilon_0&=& \cosh (\sigma_B \Delta\tau_U) f_0 + \sinh (\sigma_B \Delta\tau_U) f_1\ ,\nonumber\\
\epsilon_1&=& \sinh (\sigma_B \Delta\tau_U) f_0 + \cosh (\sigma_B \Delta\tau_U) f_1\ , \nonumber\\
\epsilon_2&=& \cos  (\sigma_R \Delta\tau_U) f_2 + \sin  (\sigma_R \Delta\tau_U) f_3\ , \nonumber\\   
\epsilon_3&=& -\sin  (\sigma_R \Delta\tau_U) f_2 + \cos  (\sigma_R \Delta\tau_U)  f_3\ ,
\end{eqnarray}
with
\beq
\label{sigBR}
\sigma_B = \sqrt{\frac{I_1+\sqrt{I_1^2+I_2^2}}{2}}\ ,\qquad
\sigma_R =  \sqrt{\frac{-I_1+\sqrt{I_1^2+I_2^2}}{2}}\ .
\eeq
and
\beq
\label{invar}
I_1=\kappa^2-(\tau_1^2+\tau_2^2) , \qquad I_2=2\kappa \tau_2 \ .
\eeq

The orthonormal frame vectors $\epsilon_\alpha$ are then parallel transported along $U$.
Note that this frame reduces to the Fermi-Walker transported frame (\ref{fwframe}) in the limit $\kappa\to0$ as expected since 
\beq
(\tilde\alpha, 
\sigma_B)\to (0,0)\ ,\
\sigma_R\to||\omega_{\rm (FS)}||\ ,\
(\cos\Theta,\sin\Theta)  \to\frac{1}{||\omega_{\rm (FS)}||} (\tau_1,-\tau_2)
\ .
\eeq

Finally if $\tau_2=0$, as it does identically in the
equatorial plane of a reflection symmetric spacetime like the Kerr spacetime,
the initial boost is all that is required, with simpler boost parameter formulas
\begin{eqnarray}
\label{eq:boost-rot}
|\kappa|>|\tau_1|:
&&\sigma_B=(\kappa^2-\tau_1^2)^{1/2}\ ,\ \sigma_R=0\ ,
\nonumber\\
&&
\tanh\tilde \alpha=\frac{\tau_1}{\kappa}\ , \
\sinh\tilde \alpha=\frac{{\rm sgn }[\kappa]\tau_1}{\sigma_B}\ , \
\cosh\tilde \alpha=\frac{|\kappa|}{\sigma_B}\ ,\
\nonumber\\
|\kappa|<|\tau_1|:&&\sigma_R=(\tau_1^2-\kappa^2)^{1/2}\ ,\ \sigma_B=0\ ,
\nonumber\\
&&
\tanh\tilde \alpha=\frac{\kappa}{\tau_1}\ , \
\sinh\tilde \alpha=\frac{{\rm sgn }[\tau_1]\kappa}{\sigma_R}\ , \
\cosh\tilde \alpha=\frac{|\tau_1|}{\sigma_R}\ ,\
\end{eqnarray}
while $f_1=E_1$ and $f_3=E_3$ are unchanged at the first step.
Analogous to the case of a null electromagnetic 2-form in which the electric and magnetic fields are orthogonal and can be transformed by an appropriate boost to a pure electric or pure magnetic field depending on which has the greater intensity, a parallel transported frame is obtained from $f_\alpha$ by a constant rate boost in the $f_0$-$f_1$ plane or a constant rate rotation in the $f_1$-$f_2$ plane, while when the intensities are equal, a null rotation results \cite{bccj}.
Eqs.~(\ref{parframe}) continue to hold when $\sigma_R=0$, $\sigma_B\neq0$,
but in the case $\sigma_R\neq0$, $\sigma_B=0$, the rotation occurs instead in the $f_1$-$f_2$ plane.

\section{Test gyroscopes and circular holonomy}

The Frenet-Serret frame is the key for finding explicit expressions for the remaining two types of physical frames given above. We now discuss each of them in the context in which they are most often used to the interpret spacetime geometry associated with circular orbits in stationary axisymmetric spacetimes.
Fermi-Walker frames describe the precession of test gyroscopes, whose spin vector is Fermi-Walker transported along their world lines and therefore has constant components in a Fermi-Walker frame. Comparison with the original symmetry adapted spatial frames associated with observers at spatial infinity then gives some idea of how strong dragging effects are in the gravitational field of the spacetime. Parallel transported frames give a more general idea of how strong such curvature effects are, and holonomy invariance for vectors undergoing parallel transport provides some natural markers for this effect.
In each case vectors which undergo the respective transport have constant components in the associated frame.

Although the notion of what constitutes a single loop of an orbit depends on one's choice of observers, it is common to compare the change in transported vectors after one revolution of the azimuthal coordinate $\phi$ corresponding to the observers considered to be nonrotating with respect to spatial infinity. This is a proper time interval of $\Delta\tau_U = 2\pi/|\Omega_U|$, where $\Omega_U$ is the proper orbital angular velocity.

\subsection{Fermi-Walker transport}

A Fermi-Walker frame $\{F_\alpha\}$ along a circular orbit can be operationally defined by a set of three torque-free test gyroscopes with orthogonal spin vector directions, i.e., each having its associated spin vector ${\mathcal S}$ in the local rest space orthogonal to the 4-velocity $U$ and undergoing Fermi-Walker transport along the orbit, namely
\beq
U\cdot {\mathcal S}=0\ , \qquad 
\nabla_{({\rm fw}, U)}{\mathcal S}=0\ ,
\eeq
implying that the adapted Fermi-Walker frame components have ${\mathcal S}^0=0$ while ${\mathcal S}^a$ are constants along the orbit.

From (\ref{fwframe}) one sees that the spin vector only rotates in the plane orthogonal to $F_3=H_3$ (the direction of the Frenet-Serret angular velocity) in the local rest space of $U$, by an angle 
$\Delta\varphi=||\omega_{\rm (FS)}||\Delta\tau_U$ governed by the Frenet-Serret angular velocity, which is distinct from the proper angular velocity $\Omega_U$ of the orbit itself which determines the change $\Delta\phi=\Omega_U \Delta\tau_U$
in the orbital azimuthal angle, leading to a relative precession of the spin vector compared with the symmetry adapted frame associated with the observers at spatial infinity. The observers at infinity are locally represented by applying the inverse orbital rotation locally to the symmetry adapted frame, but these spatial axes can be compared with Fermi-Walker axes in the local rest space of the orbit only after performing a relative boost between them. By comparing the spin vector with its initial value only after an integer number of azimuthal orbit loops, one sidesteps the local rotation of the symmetry adapted axes due to the orbital motion.

During one orbital revolution during which $|\Delta\phi|=|\Omega_U| \Delta\tau_U = 2\pi$ and returns to the same azimuthal position in the orbit, corresponding to a proper orbital period $\Delta\tau_U=2\pi/|\Omega_U|$, the spin direction changes by the angle $|\Delta\varphi| = 2\pi ||\omega_{\rm (FS)}||/|\Omega_U| $.
The difference defines the relative change in angle of the spin direction as calculated by Rindler and Perlick \cite{rind-perl}
and Iyer and Vishveshwara \cite{iyer-vish}
\beq\label{precession}
  |\Delta\varphi| - |\Delta\phi| =  2\pi (||\omega_{\rm (FS)}||/|\Omega_U|-1)\ ,
\eeq
or dividing by the proper orbital period, we get the proper angular velocity of this relative precession $||\omega_{\rm (FS)}|| - |\Omega_U|$, which turns out to be opposing the orbital motion when signs are taken into account through the relative orientation of the Fermi-Walker and symmetry adapted frames. When the spacetime is nearly flat along a circular orbit, the spin vector must rotate in the opposite direction relative to the orbital rotation by almost the same amount $||\omega_{\rm (FS)}|| \approx |\Omega_U|$ in order to almost maintain its direction with respect to nearly flat Cartesian axes. As the spacetime becomes more curved, these two angular velocities begin to differ appreciably so that a significant difference occurs during one azimuthal orbit. One way to mark the size of this difference is to note how many azimuthal orbital loops the spin vector takes to undergo exactly one or more full revolutions by $2\pi$. 

The condition that the precession angle $|\Delta\varphi|$ is some multiple $m\ge0$ of $2\pi$ after $n>0$ orbital periods 
$\Delta\tau_U = 2 n \pi/|\Omega_U|$
so that the spin vector boosted back to the ZAMO local rest space has the same orientation with respect to the symmetry adapted frame as it initially had is correspondingly
\beq
\label{spin_hol}
  2 m \pi = 2 n \pi ||\omega_{\rm (FS)}||/|\Omega_U| \qquad \rightarrow \qquad 
\frac{||\omega_{\rm (FS)}||}{|\Omega_U|}=\frac{m}{n}\ ,
\eeq
which is a condition on the angular velocity of the orbit which may or may not be satisfied at some circular orbit position for some values of these nonnegative integers. 
The values $(m,n)=(1,1)$ characterize the exact compensation of the orbital rotation by the rotation of the boosted spin vector, so that the boosted spin vector has constant components in the frame (\ref{nonrotframeinfinity}), provided that they are in opposing directions.
If $||\omega_{\rm (FS)}||=0$ for $|\Omega_U|\neq0$, then the boosted spin vector has constant components in the spherical frame, and hence undergoes one complete revolution forwards during one azimuthal loop with respect to that same frame (\ref{nonrotframeinfinity}), i.e., in the same direction as the orbit, equivalent to $(m,n)=(0,1)$.  On the other hand if $(m,n)=(2,1)$, the  boosted spin vector will rotate backwards one complete revolution with respect to that frame. Thus $m-1=1,0,-1$ with $n=1$ corresponds to one revolution forwards, no revolution, and one revolution backwards with respect to the frame not rotating with respect to infinity. Higher $m$ values would lead to larger rotations, and similarly the same $m$ values with $n=2$ would lead to half as strong a rotation and so on. 
However, one needs a concrete example to see what relevance this general discussion has for interpreting actual spacetimes, in particular how the plane of the rotation compares to the ``orbital plane" apart from the boost between local rest spaces.

\subsection{Parallel transport}

The corresponding discussion for parallel transport is more complicated by the additional boost in the azimuthal direction relative to the local rest space of the circular orbit, but it is a more conventional differential geometry topic: holonomy, namely how vectors are changed under parallel transport around closed loops in a manifold with a connection.
Rothman et al \cite{roth} first broached this subject for spatial holonomy alone by showing that a generic vector which is parallel-transported around a spacelike equatorial circular orbit following the closed azimuthal coordinate line in the Schwarzschild spacetime is rotated with respect to its initial orientation with respect to a symmetry adapted frame after a closed azimuthal loop. This holonomy rotation is the identity after $n$ circuits of such a closed spatial loop at radius $r$ if $n$ and $r$ satisfy an appropriate condition. 
The modification of this holonomy invariance for rotating black holes was investigated by Maartens et al \cite{maartens}, who discovered that for the parallel transport of a vector field around a corresponding equatorial circular orbit in the Kerr spacetime 
%as well 
in addition to a net rotation, a boost also occurred relative to the initial vector, concluding that rotation of the source and its associated gravitomagnetism leads to this more general holonomy. They extended this discussion to the larger class of stationary axisymmetric spacetimes that are reflection symmetric about their equatorial planes. 

Bini et al \cite{bjm_chce,bcj,bccj,bjnc} generalized this discussion of holonomy associated with circular orbits to all spacetimes which admit clock effects (i.e., allow oppositely directed circular geodesics within the same axially symmetric cylinderical symmetry group orbits, so clocks can be compared at intersection points, and hence spin vectors as well) and to circular orbits of any causality. This allows one to form closed loops in spacetime from a pair of oppositely rotating timelike circular orbits, and thus compare vectors at invariantly defined intersection points of these orbits. Thus in addition to examining the effects of spacetime rotation on the proper times between meeting points, one can also compare both parallel transported and spin vectors which were initially aligned, apart from the boost between local rest spaces.

If $X$ is a generic vector undergoing parallel transport along $U$, its components in the parallel transported frame $\epsilon_\alpha $ are all constant 
\beq
\label{Xpartrasp}
X=X^\alpha(0)\epsilon_\alpha\ \ .
\eeq
Invariance for the vector (\ref{Xpartrasp}) after a single orbital circuit of $\Delta\phi=2\pi$
can occur in the general case only if it does not undergo a boost, which happens only when it is orthogonal to the boost plane, i.e., if $X^0(0)=0=X^3(0)$. In this case $\sigma_R$ takes the place of the Frenet-Serret angular velocity $||\omega_{(FS)}||$ of the previous discussion, and the invariance condition is then
\beq\label{holonomycondition}
\frac{\sigma_R}{|\Omega_U|}=\frac{m}{n}\ ,
\eeq
now referred to as ``holonomy invariance," although strictly speaking a circular orbit alone does not define a closed loop in spacetime.

\section{Equatorial plane orbits}

To make this discussion more concrete it is helpful to specialize to the case of circular motion in the equatorial plane of a reflection-symmetric stationary axisymmetric spacetime, and then to the explicit such situation in the Kerr and Schwarzschild spacetimes. The additional reflection symmetry causes the second torsion to vanish identically ($\tau_2=0$), simplifying matters considerably.

\subsection{Fermi-Walker transport}

The Fermi-Walker spatial triad (\ref{fwframe}) under the condition $\tau_2=0$ becomes
\begin{eqnarray}
\label{fwframeeq}
F_1 &=& \cos(\tau_1 \Delta\tau_U) E_1 - \sin(\tau_1 \Delta\tau_U) E_2 \nonumber \\
F_2 &=& \sin(\tau_1 \Delta\tau_U) E_1 + \cos(\tau_1 \Delta\tau_U)E_2 \nonumber \\
F_3 &=& E_3\ .
\end{eqnarray}
The basis vectors simply rotate with frequency $\tau_1$ in the plane of the acceleration vector and its covariant derivative.

\subsection{Parallel transport}

The vanishing of the second torsion $\tau_2$ means that $E_3$ is parallel transported, so that the orientation of the plane containing the acceleration vector and its covariant derivative does not rotate along the orbit with respect to parallel transport.
The quantities $\sigma_B$ and $\sigma_R$ defined by Eq.~(\ref{sigBR}) characterizing the constant boost and rotation rates of the parallel propagated frame reduce to 
\beq\label{sigmaBRequat}
\sigma_B = \sqrt{(I_1+|I_1|)/2}\ , \qquad 
\sigma_R = \sqrt{(-I_1+|I_1|)/2}\ , 
\eeq
since the invariants (\ref{invar}) simplify to
$I_1=\kappa^2-\tau_1^2$, $I_2=0$.

One must distinguish following three cases:
 
\begin{enumerate}

\item  $|\tau_1|<|\kappa|$: $I_1>0$, $\sigma_B=\sqrt{\kappa^2-\tau_1^2}$, $\sigma_R=0$ (boost dominated).

Using Eqs. (\ref{eq:boost-rot}) the constant boost described above then takes the simpler form
\beq
f_0={\rm sgn}[\kappa]\sigma_B^{-1}(\kappa E_0 +\tau_1 E_2)\ , \qquad  
f_2={\rm sgn}[\kappa]\sigma_B^{-1}(\tau_1 E_0+ \kappa E_2)\ ,
\eeq
corresponding to $\tanh\tilde\alpha=\tau_1/\kappa$, with $f_1=E_1$, $f_3=E_3$.
The parallel transported frame is then obtained by a constant rate boost
\begin{eqnarray}
\epsilon_0 &=& \cosh (\sigma_B \Delta\tau_U) f_0 - \sinh (\sigma_B \Delta\tau_U) f_1\ ,\nonumber\\
\epsilon_1 &=& -\sinh (\sigma_B \Delta\tau_U) f_0 + \cosh (\sigma_B \Delta\tau_U) f_1\ ,
\end{eqnarray}
with $\epsilon_2= f_2$, $\epsilon_3= f_3$ unchanged.
\\

\item  $|\tau_1|>|\kappa|$:  $I_1<0$, $\sigma_B=0$, $\sigma_R=\sqrt{\tau_1^2-\kappa^2}$ (rotation dominated).

Using Eqs. (\ref{eq:boost-rot}) the constant boost described above then takes the simpler form
\beq
f_0={\rm sgn}[\tau_1]\sigma_R^{-1}(\tau_1 E_0+ \kappa E_2)\ , \qquad  
f_2={\rm sgn}[\tau_1]\sigma_R^{-1}(\kappa E_0 +\tau_1 E_2)\ ,
\eeq
corresponding to $\coth\tilde\alpha=\tau_1/\kappa$, with $f_1=E_1$, $f_3=E_3$.
The parallel transported frame is then similarly obtained by a constant rate rotation
\begin{eqnarray}
\epsilon_1&=&\cos (\sigma_R \Delta\tau_U) f_1-\sin (\sigma_R \Delta\tau_U) f_2\ , \nonumber\\
\epsilon_2&=&\sin(\sigma_R \Delta\tau_U) f_1+\cos (\sigma_R \Delta\tau_U) f_2\ .
\end{eqnarray}
with $\epsilon_0=f_0$ and $\epsilon_3=f_3$ unchanged.
\\

\item  $\tau_1=\pm \kappa$, $I_1=0$, $\sigma_B=0=\sigma_R$ (null rotation).

Introduce the two null vectors $f_\pm =(E_0 \pm E_2)/\sqrt{2}$, keeping
$f_1=E_2$, $f_3-E_3$, leading to a convenient associated null frame.
When $\tau_1=\kappa$ the frame covariant derivatives along the orbit are then
\beq
\frac{Df_+}{d\tau_U}=0\ , \qquad  
\frac{Df_1}{d\tau_U}=\sqrt{2}\kappa f_+\ ,\qquad
\frac{Df_-}{d\tau_U}=\sqrt{2} \kappa f_1\ ,
\eeq
which shows that the new frame undergoes a null rotation in the hyperplane orthogonal to $f_3$ relative to parallel transport. Removing this null rotation leads to a parallel transported null frame, from which one can obtain the associated orthonormal frame if desired
\beq
\epsilon_1=  f_1 -\sqrt{2}\kappa f_+ \Delta\tau_U , \qquad
\epsilon_-= f_- +\kappa^2 f_+ \Delta\tau_U^2-\sqrt{2}\kappa f_1 \Delta\tau_U\ ,
\eeq
keeping $\epsilon_+ = f_+$, $\epsilon_3=f_3$ unchanged.
The case $\tau_1=-\kappa$ is analogous and the result is exactly the same exchanging the roles of $f_+$ and $f_-$.
\end{enumerate}

Note that only in the rotation dominated case 2 is it possible to have periodic motion allowing vectors confined to the plane of the rotation to return to their original orientations after an integral number of azimuthal loops.

\section{Kerr spacetime}

The Kerr spacetime is the most interesting member of this symmetry class for explicit evaluation of this mathematical structure, and its equatorial plane connects most closely with our Newtonian experience with circular orbits in a gravitational field. 
In standard Boyer-Lindquist coordinates its metric is
\beq
\rmd s^2 = -\left(1-\frac{2Mr}{\Sigma}\right)\rmd t^2 -\frac{4aMr}{\Sigma}\sin^2\theta\rmd t\rmd\phi+ \frac{\Sigma}{\Delta}\rmd r^2 +\Sigma\rmd \theta^2+\frac{\Lambda}{\Sigma}\sin^2 \theta \rmd \phi^2\ ,
\eeq
where $\Delta=r^2-2Mr+a^2$, $\Sigma=r^2+a^2\cos^2\theta$ and $\Lambda = (r^2+a^2)^2-\Delta a^2\sin^2 \theta$. Here $a$ and $M$ are the specific angular momentum and total mass characterizing the spacetime. The event horizons are located at $r_\pm=M\pm\sqrt{M^2-a^2}$. 

The ZAMO family of fiducial observers exists everywhere outside the outer horizon $r_+$.
The lapse function and shift component function are
\beq
(N,\,N^\phi) 
= (
   \left[\Delta\Sigma/\Lambda\right]^{1/2}\ ,\,-2aMr/\Lambda
  )\ .
\eeq
In the equatorial plane $\theta=\pi/2$ the only nonvanishing components of the ZAMO kinematical quantities are radial
\begin{eqnarray}
\label{zamoequatkin}
\A_{\hat r} & = &
\frac{M \Delta^{-1/2}\left[(r^2+a^2)^2-4 a^2 M r\right]}
{r^2 (r^3+a^2 r+2 M a^2)}
\ ,\nonumber\\
{\theta_{\hat \phi}}(n)_{\hat r} 
& = &
\frac{M a (3 r^2+a^2)}{r^2 (r^3+a^2 r+2 M a^2)}
\ ,\nonumber\\
{k_{(\rm lie)}}(n)_{\hat r} 
&=&
-\frac{(r^3-a^2M)\sqrt{\Delta}}{r^2(r^3+a^2r+2a^2M)}\ .
\end{eqnarray}

There exists a whole collection of geometrically special circular orbits  \cite{idcf1,idcf2,bjm,bjdf} in the equatorial plane, including first and foremost the co-rotating $(+)$ and counter-rotating $(-)$ timelike circular geodesics whose angular and linear velocities are respectively
\beq
\zeta_{({\rm geo})\, \pm}\equiv\zeta_{\pm}
=\left[a\pm (M/r^3)^{1/2}\right]^{-1}\ , \
\nu_{({\rm geo})\, \pm}\equiv \nu_\pm 
=\frac{a^2\mp2a\sqrt{Mr}+r^2}{\sqrt{\Delta}(a\pm r\sqrt{r/M})}\ .
\eeq 
The corresponding timelike conditions $|\nu_\pm|<1$ identify the allowed regions $r>r_{{(\rm geo)}\pm}$ for the radial coordinate where co/counter-rotating geodesics exist, where the null circular orbits occur at 
\beq
r_{{(\rm geo)}\pm}
=2M\left\{1+\cos\left[\frac23\arccos\left(\pm\frac{a}{M}\right)\right]\right\}\ .
\eeq
Closely related to these are the ``geodesic meeting point observers'' defined by their alternating successive intersection points \cite{idcf2}, with velocity
\beq
\nu_{\rm (gmp)}=\frac{
\nu_{+}+\nu_{-}}{2}=-\frac{aM(3r^2+a^2)}{\sqrt{\Delta}(r^3-a^2M)}\ .
\eeq
The linear velocities characterizing these special orbits are related to the ZAMO kinematical quantities (\ref{zamoequatkin}) by
\beq
\A_{\hat r} = {k_{(\rm lie)}}(n)_{\hat r}\, \nu_+\nu_-\ , \qquad
{\theta_{\hat \phi}}(n)_{\hat r}
=-k_{\rm(lie)}(n)_{\hat r} \, \nu_{\rm (gmp)}\ .
\eeq
We now have expressions for all the geometrical and kinematical quantities which are needed to specify the Frenet-Serret, Fermi-Walker and parallel propagated frames.

The Frenet-Serret frame is given by Eq.~(\ref{FSframe}) with $\chi=0$, namely
\begin{eqnarray}
\label{FSframeeq}
E_0 &=&\cosh \alpha\, n+\sinh \alpha\, e_{\hat \phi} \ ,\qquad
E_1=e_{\hat r}\ ,\ 
\nonumber \\
E_2&=&\sinh \alpha\, n +\cosh \alpha\, e_{\hat \phi}\ ,\qquad 
E_3=- e_{\hat \theta}\ ,
\end{eqnarray}
recalling that $\nu=\tanh\alpha,\gamma=\cosh\alpha$.
The second torsion $\tau_2$ vanishes, while the geodesic curvature $\kappa$ and the first torsion $\tau_1$ are given by
\begin{eqnarray}
\label{ketau1}
\kappa &=&{k_{\rm (lie)}}(n)_{\hat r}\gamma^2 (\nu-\nu_+)(\nu-\nu_-)\ , \nonumber \\
\tau_1&=& k_{\rm(lie)}(n){}_{\hat r}\nu_{\rm (gmp)}\gamma^2 (\nu-\nu_{{\rm (crit)}+})(\nu-\nu_{{\rm (crit)}-})\ ,
\end{eqnarray}
where the first torsion vanishes for the ``critically accelerated orbits'' \cite{fdfacc,semerak,idcf2}
\begin{eqnarray}
\nu_{{\rm (crit)}\pm}&=&\frac{\gamma_- \nu_- \mp \gamma_+ \nu_+}{\gamma_- \mp \gamma_+} \nonumber \\
&=& -\frac1{2Ma(3r^2+a^2)\sqrt{\Delta}}\Big[-2a^2M(a^2-3Mr)+r^2(r^2+a^2)(r-3M)\nonumber\\
&&\pm(r^3+a^2r+2a^2M)\sqrt{r}\sqrt{r(r-3M)^2-4a^2M}\Big]\ .
\end{eqnarray}
$|\nu_{{\rm (crit)}-}|<1$ in both regions $r_+<r<r_{{(\rm geo)}+}$ and $r>r_{{(\rm geo)}-}$ where $|\nu_{{\rm (crit)}+}|>1$, while $|\nu_{{\rm (crit)}+}|<1$ holds for the complementary region $r_{{(\rm geo)}+}<r<r_{{(\rm geo)}-}$. The timelike such orbits define extremely accelerated observers.

The Fermi-Walker spatial frame is given by Eq. (\ref{fwframeeq}) with Frenet-Serret triad (\ref{FSframeeq}) and first torsion given by Eq. (\ref{ketau1}).
This frame simply rotates in the boost to the local rest space of the orbit of the $r$-$\phi$ plane tangent to the equatorial plane to which the orbit is confined, describing a simple rotation of spin vectors along the orbit, modulo the relative boost needed for comparison with the symmetry adapted axes. The net precession after $n$ azimuthal loops results from the comparison of the proper angular velocity $\tau_1$ of this rotation with the proper orbital velocity
\beq
\label{omUkerr}
\Omega_U=\gamma\frac{2aM+\nu r\sqrt{\Delta}}{\sqrt{r\Delta}\sqrt{r^3+a^2r+2a^2M}}=\frac{\gamma}{\sqrt{g_{\phi\phi}}}(\nu-\nu_0)\ . 
\eeq
The spin vector in the boosted $r$-$\phi$ plane which returns to its original position after $n$ azimuthal loops satisfies 
\beq
 2\pi  n\frac{\tau_1}{|\Omega_U|}= 2\pi m\ , 
\eeq
according to Eq.~(\ref{spin_hol}), where here we allow $m$ to change sign with $\tau_1$.

\begin{figure}
\typeout{*** EPS figure 1}
\begin{center}
\includegraphics[scale=0.40]{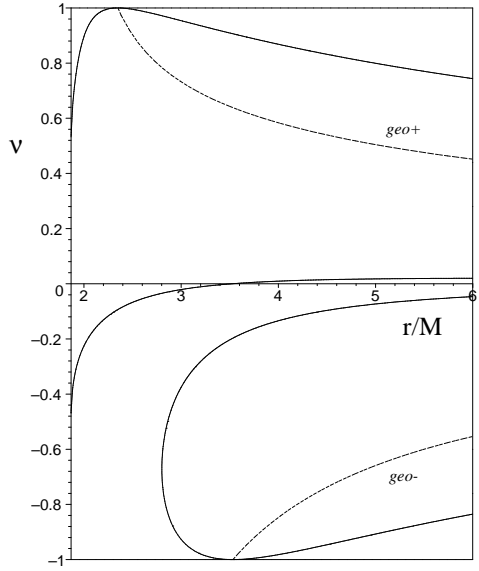}
\end{center}
\caption{The linear velocities $\nu^{\rm S}_{(m,n)}$ satisfying the spin invariance condition (\ref{spin_hol}) with $m=1$ and $n=1$ are plotted as functions of the radial parameter $r/M$ in the Kerr case for $a/M=0.5$. For this $\nu$  an initial vector rotates by $\pm 2\pi$ under Fermi-Walker transport after $n=1$ azimuthal loop. 
}
\label{fig:1}
\end{figure}

\begin{figure} 
\typeout{*** EPS figure 2}
\begin{center}
$\begin{array}{cccc}
\includegraphics[scale=0.4]{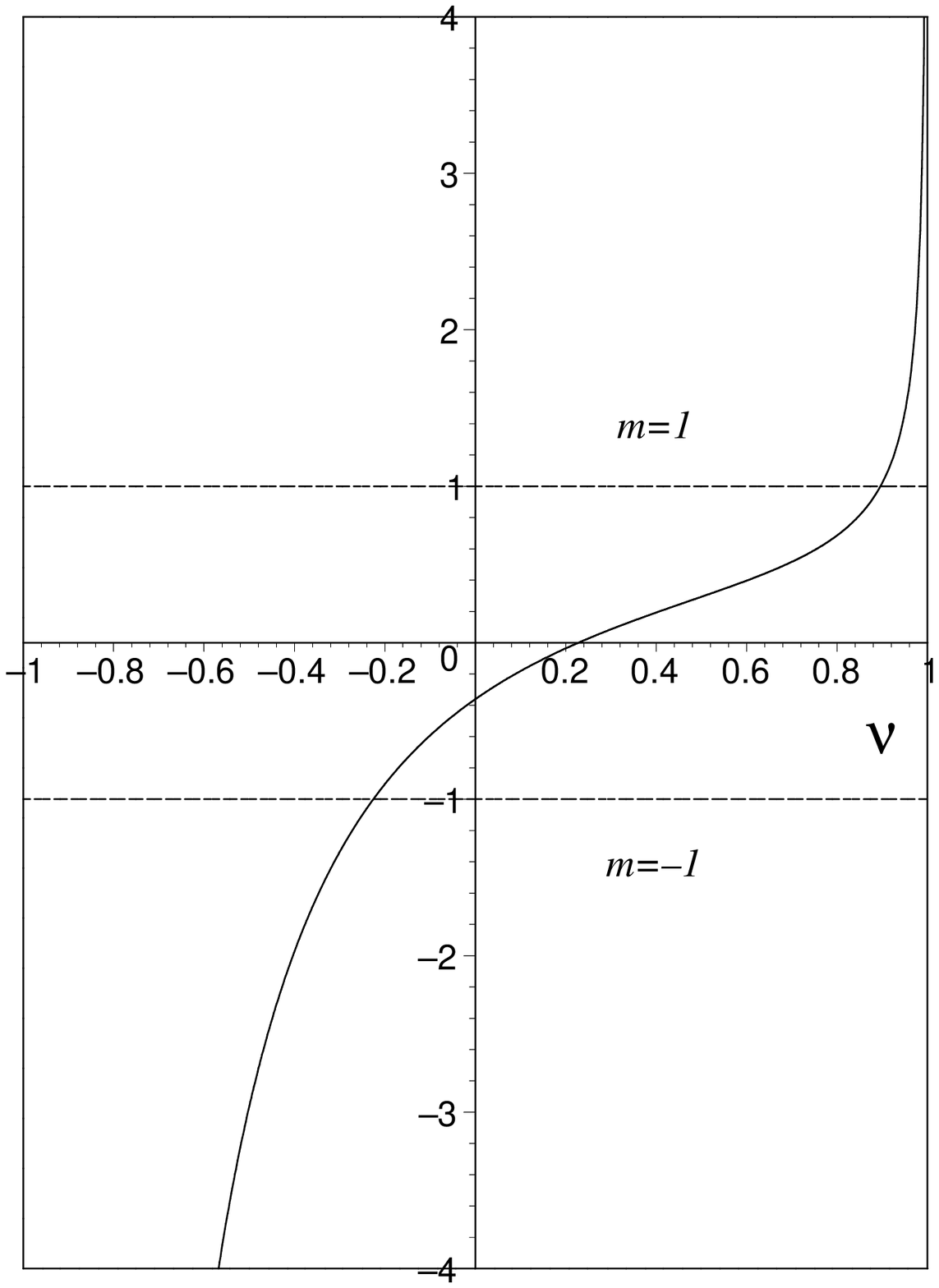}&\qquad
\includegraphics[scale=0.4]{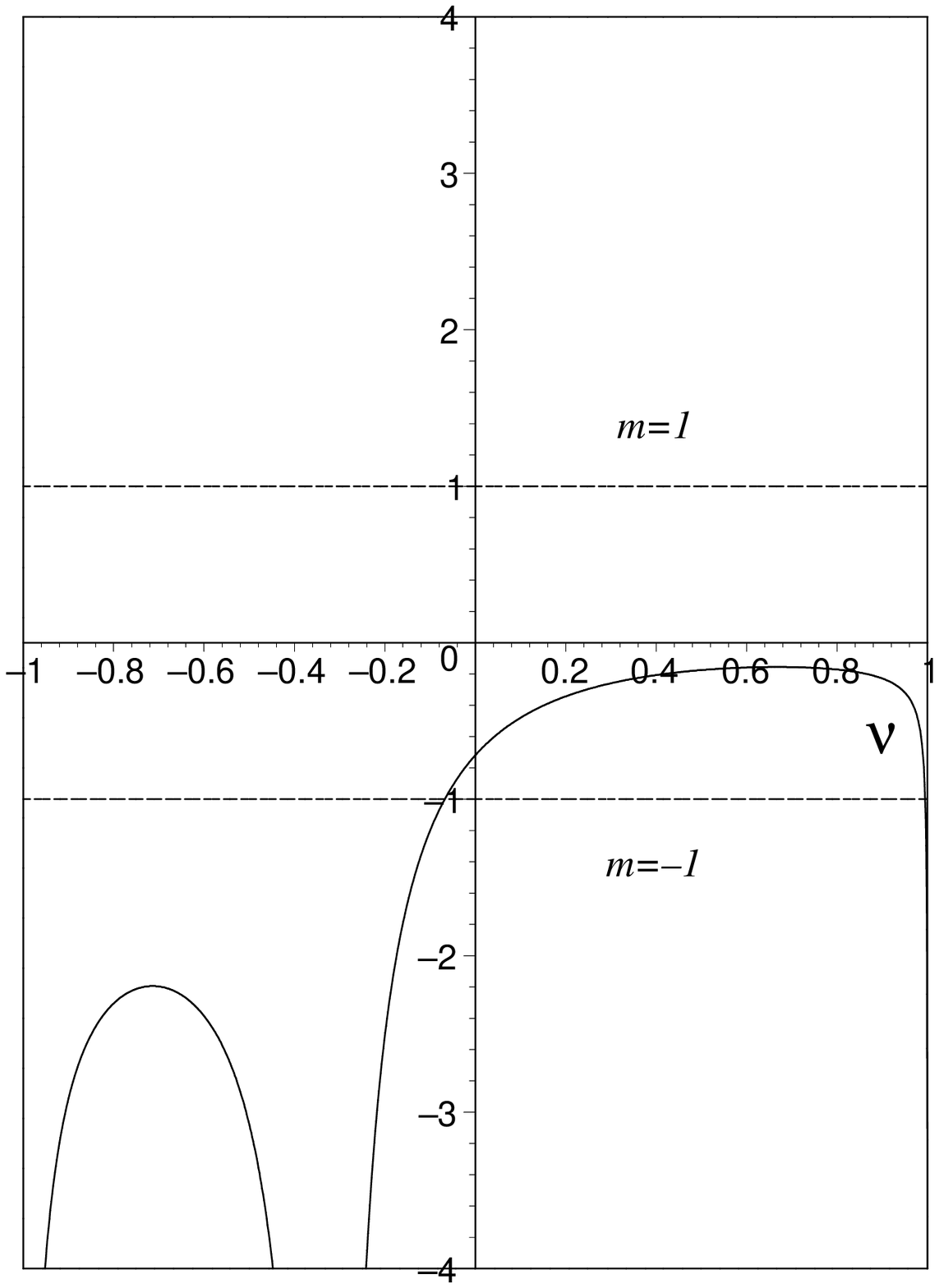}&\\[.2cm]
\mbox{(a)} &\qquad \mbox{(b)}&\\[.6cm]
\includegraphics[scale=0.4]{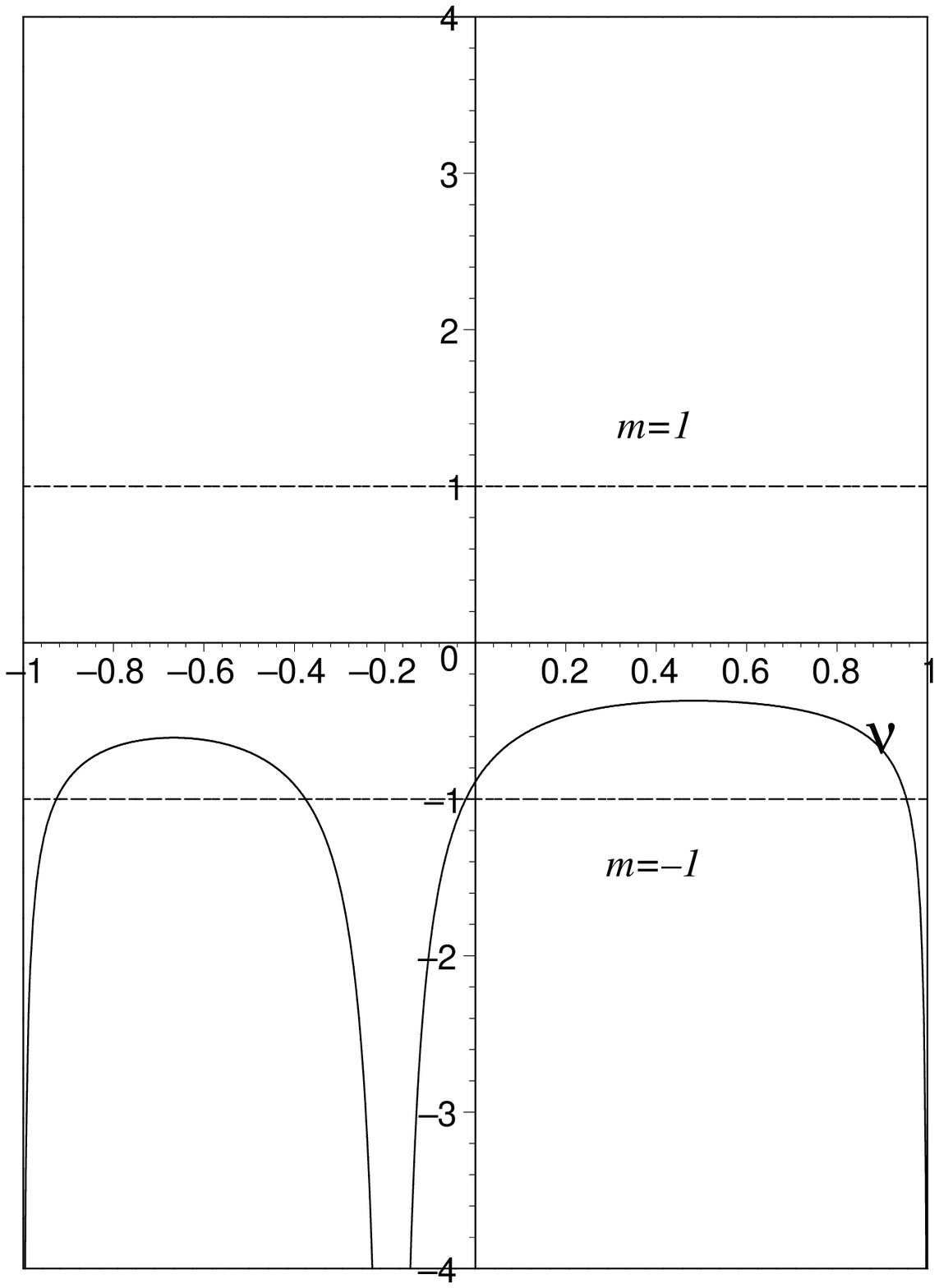}&\qquad
\includegraphics[scale=0.4]{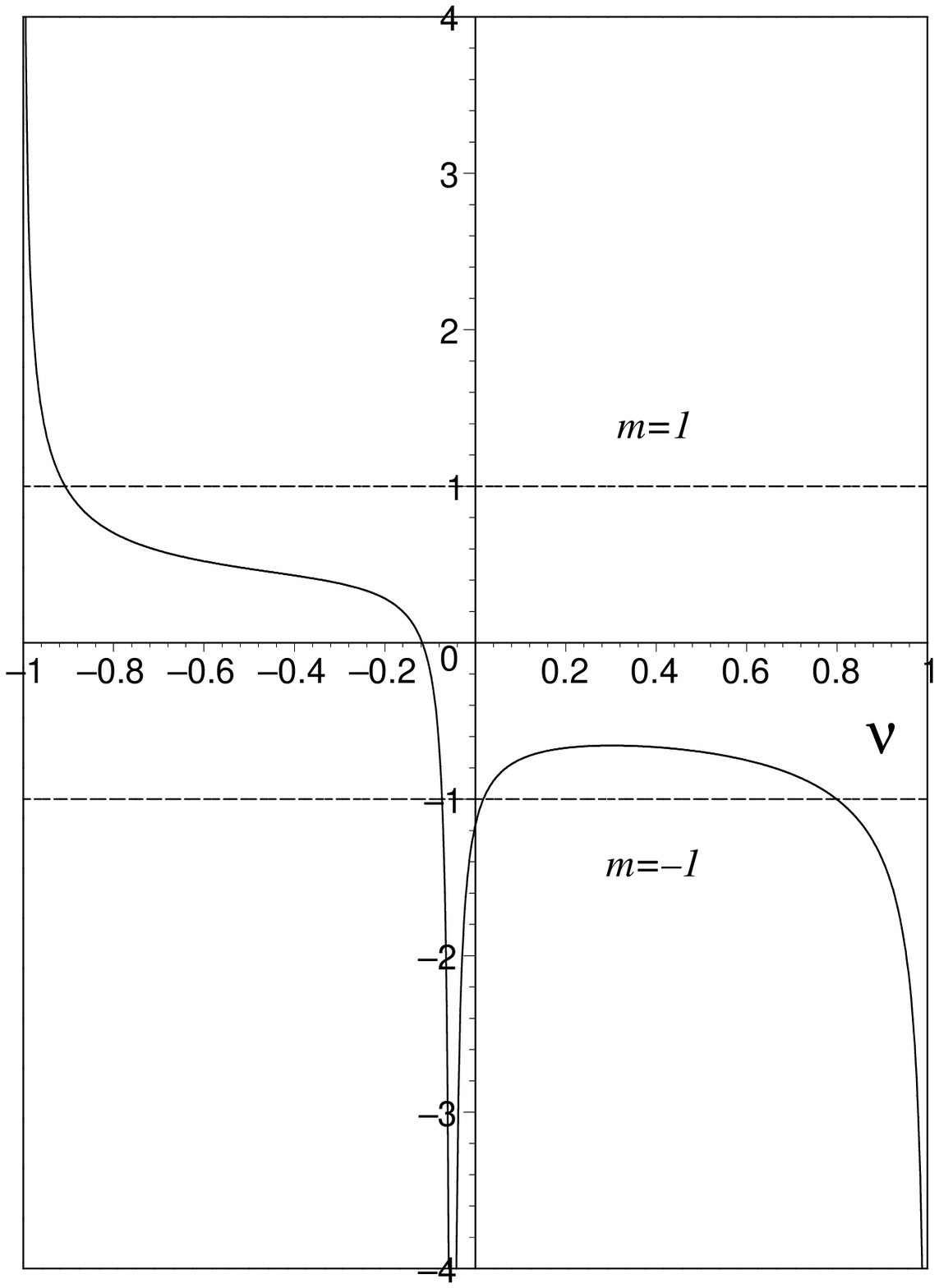}&\\[.2cm]
\mbox{(c)} &\qquad \mbox{(d)}
\end{array}
$
\end{center}
\caption{The behavior of $\tau_1/|\Omega_U|$ as a function of $\nu$ is shown for fixed values of the radial coordinate $r/M=[2,2.5,3,5]$ (from (a) to (d) respectively), with $a/M=0.5$. Horizontal dashed lines represent the constant values $m=\pm1$. The intersection points give the values of velocities corresponding to vanishing spin precession after $n=1$ azimuthal loop.
The intersections of the curves with the horizontal axis give instead the velocities of extremely accelerated observers corresponding to the case $m=0$. 
}
\label{fig:2}
\end{figure}

The corresponding corotating and counter-rotating velocities $\nu^{\rm S}_{(m,n)}$ for which this condition is satisfied are shown in Fig.~\ref{fig:1} as functions of the radial coordinate for 
$m=1$, $n=1$ 
to give a flavor of how this compares to the geodesic velocities.
Fig.~\ref{fig:2} shows instead the behavior of the quantity $\tau_1/|\Omega_U|$ as a function of $\nu$ for fixed values of the radial coordinate: when the curves reach the integer values $m=\pm1, \ldots$, we see the values of velocities corresponding to spin vectors returning to their original positions after $n=1$ azimuthal loop. A vertical asymptote exists at the static observer velocity where $\Omega_U=0$; nearby orbits rotate so slowly that small precession effects accumulate during the long period, explaining the $m=-1$ velocities close to the location of the asymptote. The high velocity crossing points are basically due to the special relativistic Thomas precession where sufficiently high $\gamma$ leads to large precession. $\tau_1$ vanishes for the extremely accelerated observers, corresponding to the $m=0$ case.

% figure 1

% figure 2

\begin{figure}
\typeout{*** EPS figure 3}
\begin{center}
\includegraphics[scale=0.40]{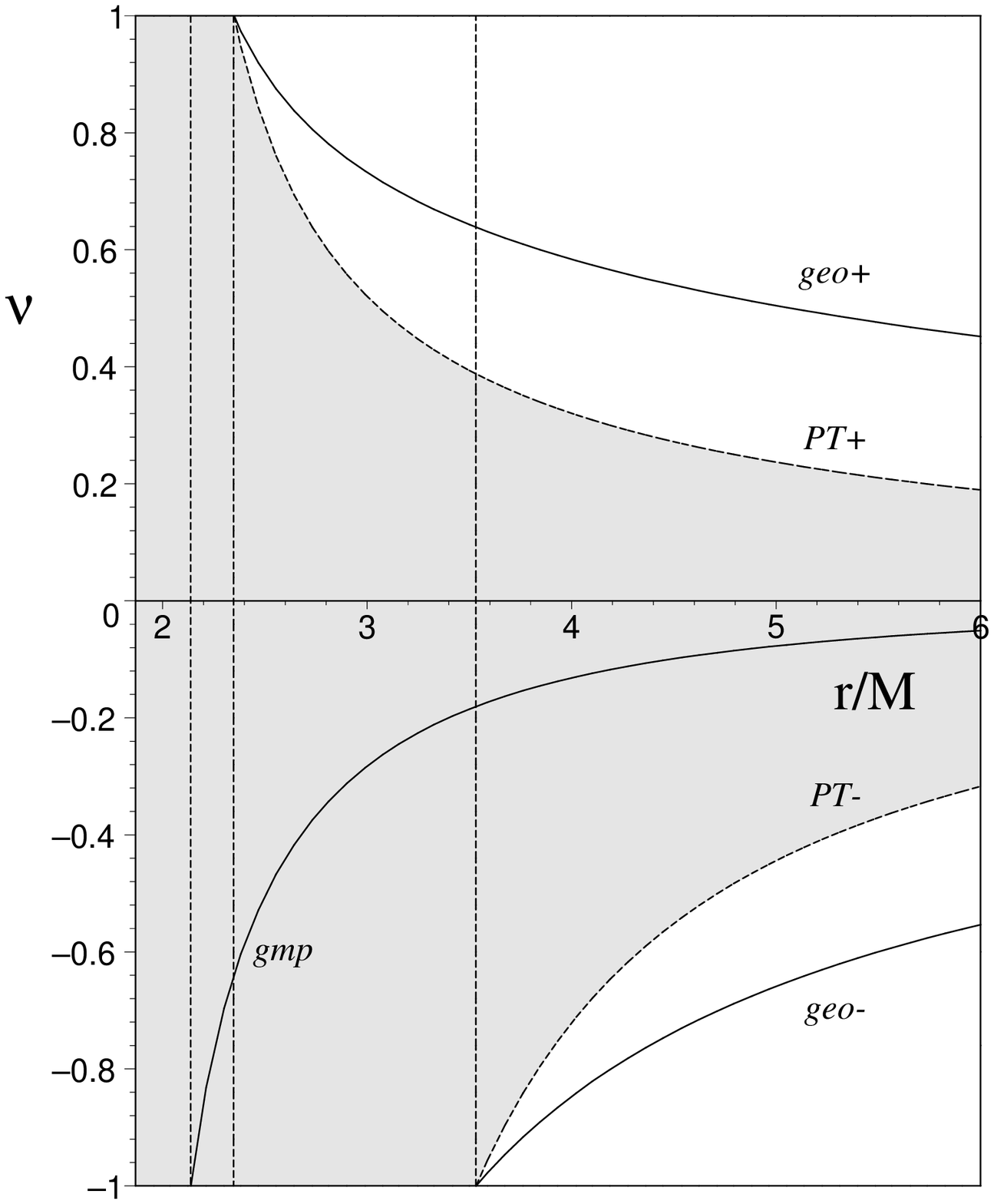}
\end{center}
\caption{The relative velocities of geodesics ($\nu_\pm$), of geodesic meeting point observers ($\nu_{\rm (gmp)}$) and those of parallel-transported orbits $(\nu_{({\rm PT} \pm)})$  are plotted a functions of $r/M$ in the Kerr case for $a/M=0.5$. The shaded region corresponds to the  boost dominated case.
}
\label{fig:3}
\end{figure}

\begin{figure}
\typeout{*** EPS figure 4}
\begin{center}
\includegraphics[scale=0.40]{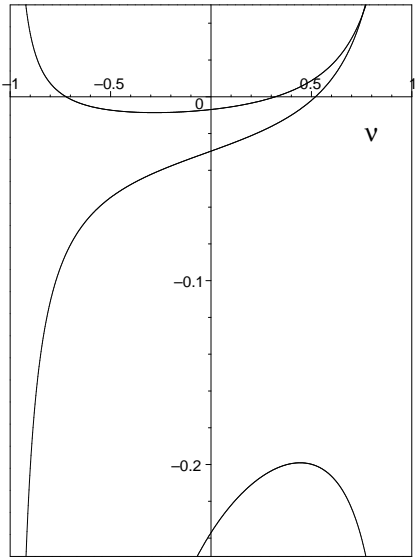}
\end{center}
\caption{The quantity $\tau_1^2-\kappa^2$ is plotted as a function of $\nu$ in the Kerr case for $a/M=0.5$ for fixed values of $r/M=2.2,3,4$ (from bottom to top). Positive values 
correspond to the rotation dominated case, which is excluded for values of $r$ in between the outer horizon and the corotating geodesic radius, in agreement with Fig. \ref{fig:3}.
}
\label{fig:4}
\end{figure}

Parallel transport in the equatorial plane was discussed extensively in \cite{bjm_chce}.
Starting from Eq.~(\ref{ketau1}) we have
\beq
\label{tau12menok2}
\tau_1^2-\kappa^2=k_{\rm(lie)}(n){}_{\hat r}^2 \frac{\gamma^2}{\gamma_{\rm (gmp)}^2} (\nu-\nu_{{(\rm PT} +)} )(\nu-\nu_{{(\rm PT} -)})
\eeq
where
\beq
\nu_{({\rm PT} \pm)}=\frac{\nu_{\rm (gmp)}\mp \nu_+\nu_-}{1\mp \nu_{\rm (gmp)}},
\eeq
have been introduced in \cite{bccj}.
Notice that the values $\nu_{({\rm PT} \pm)}=\pm 1$ occur when respectively $\nu_\pm=\pm 1$.
A rotation dominated case ($|\tau_1|>|\kappa|$) occurs for all orbits  with $\nu < \nu_{{(\rm PT} -)}$ and $\nu >\nu_{{(\rm PT} +)}$;
orbits with $\nu = \nu_{{(\rm PT} \pm)}$ correspond to null rotations; finally, boost dominated cases ($|\tau_1|<|\kappa|$) correspond to the remaining allowed values of $\nu$.
The situation is shown in Fig. \ref{fig:3} with the boost dominated case indicated by shading, as well as in Fig. \ref{fig:4} where  the quantity $\tau_1^2-\kappa^2$ in Eq.~(\ref{tau12menok2}) distinguishing between boost and rotation dominance is plotted as a function of $\nu$, for fixed values of $r/M$.

\begin{figure}
\typeout{*** EPS figure 5}
\begin{center}
\includegraphics[scale=0.40]{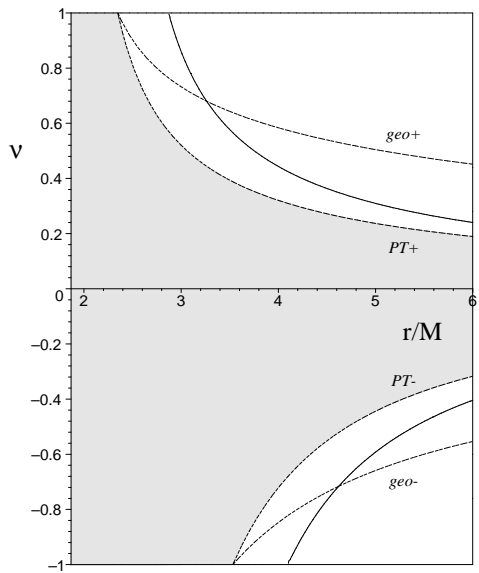}
\end{center}
\caption{The linear velocities $\nu^{\rm P}_{(m,n)}$ satisfying the holonomy invariance condition (\ref{holonomycondition}) with $m=1$ and $n=2$ are plotted as functions of the radial parameter $r/M$ in the Kerr case for $a/M=0.5$. For these orbits  an initial vector rotates by $\pm 2\pi$ under parallel transport after $n=2$ azimuthal loops. The shaded region corresponds to the  boost dominated case and the geodesics are also plotted for convenience.
}
\label{fig:5}
\end{figure}

The  frame to which the parallel propagated frame reduces in the case of equatorial orbits can be found in Section 5.2.
The quantity $\sigma_R/|\Omega_U|$  entering the holonomy condition (\ref{holonomycondition}) 
turns out to be given by
\beq
\frac{\sigma_R}{|\Omega_U|}=\frac{\sqrt{g_{\phi\phi}}|k_{\rm(lie)}(n){}_{\hat r}|}{|\nu-\nu_0|}\frac{1}{\gamma_{\rm (gmp)}}|(\nu-\nu_{{(\rm PT} +)})(\nu-\nu_{{(\rm PT} -)})|^{1/2},
\eeq
using Eq. (\ref{tau12menok2}). This ratio is strictly less than one, so comparing it with the ratio $m/n$ of the holonomy invariance condition (\ref{holonomycondition}), one must have $|m|<|n|$, i.e., in one azimuthal revolution, a parallel transported vector can only rotate less than one revolution with respect to the spherical frame.
The linear velocities $\nu^{\rm P}_{(m,n)}$ satisfying the holonomy invariance condition are shown in Fig.~\ref{fig:5}  for the largest ratio of integers case $m=1$ and $n=2$ as an example.
The expression for $\nu^{\rm P}_{(m,n)}$ is the following
\begin{eqnarray} 
\nu_{(m,n)}^{\rm P}&=&\frac{n^2K^2(\nu_{{(\rm PT} +)}+\nu_{{(\rm PT} -)})-2m^2\nu_0\pm nK\Lambda}{2(n^2K^2-m^2)}\ , \nonumber\\
\Lambda&=&\left[n^2K^2(\nu_{{(\rm PT} +)}-\nu_{{(\rm PT} -)})^2+4m^2(\nu_0-\nu_{{(\rm PT} +)})(\nu_0-\nu_{{(\rm PT} -)})\right]^{1/2}\ ,\nonumber\\
K&=&\frac{\sqrt{g_{\phi\phi}}}{\gamma_{\rm (gmp)}}|k_{\rm(lie)}(n){}_{\hat r}|\ . 
\end{eqnarray} 
It is easy to show that when $m=n$, the expression $\nu_{(m,m)}^{\rm P}$ is complex.
 
% figure 3

% figure 4

% figure 5

\subsection{The Schwarzschild spacetime limit}

The limit of vanishing rotation parameter $a=0$ for Kerr metric is the Schwarzschild metric, with lapse $N=\sqrt{-g_{tt}}$ and zero shift $N^{\phi}=0$
\beq
\label{metric}
\rmd  s^2 = -\left(1-\frac{2M}r\right)\rmd t^2 
   + \left(1-\frac{2M}r\right)^{-1} \rmd r^2 
   + r^2 (\rmd \theta^2 +\sin^2 \theta \rmd \phi^2)\ .
\eeq
In this limit the ZAMOs coincide with the static observers, and their kinematical quantities (\ref{zamoequatkin}) reduce to 
\beq
\A_{\hat r} = \frac{M}{r^2}\left(1-\frac{2M}{r}\right)^{-1/2}, \
{\theta_{\hat \phi}}(n)_{\hat r} = 0\ , \
{k_{(\rm lie)}}(n)_{\hat r} =-\frac1r\left(1-\frac{2M}{r}\right)^{1/2}\ .
\eeq

The Frenet-Serret frame for circular orbits in the equatorial plane $\theta=\pi/2$ is given by Eq.~(\ref{FSframeeq}) with $\tau_2=0$
\begin{eqnarray}
\label{kappatau1schw}
\kappa &=& {k_{\rm (lie)}}(n)_{\hat r}\gamma^2(\nu^2-\nu_K^2)
 =-\frac{\zeta_K}{\nu_K}\gamma^2(\nu^2-\nu_K^2)\ , \nonumber\\
\tau_1 &=&-{k_{\rm (lie)}}(n)_{\hat r}\frac{\gamma^2}{\gamma_K^2}\nu
= \frac{\zeta_K}{\nu_K}\frac{\gamma^2}{\gamma_K^2}\nu\ , 
\end{eqnarray}
where the quantities 
\beq
\label{geosschw}
\nu_K=\left[\frac{M}{r-2M}\right]^{1/2}\ , \ 
\gamma_K=\left[\frac{r-2M}{r-3M}\right]^{1/2}\  , \
\zeta_K=\left(\frac{M}{r^3}\right)^{1/2}\ 
\eeq
refer to the timelike equatorial circular geodesics, with $\nu_\pm=\pm\nu_K$. The boost zone interval of velocities is 
$[\nu_{(PT-)},\nu_{(PT+)}]=[-\nu_K^{1/2},\nu_K^{1/2}]$, which is inside the interval $[-\nu_K,\nu_K]$, outside of which one has pure rotations in an azimuthal plane boosted by the gamma factor $\tanh\bar\alpha = \tau_1/\kappa$.

The Fermi-Walker spatial frame is given by Eq.~(\ref{fwframeeq}) with Frenet-Serret triad (\ref{FSframeeq}) and first torsion given by Eq.~(\ref{kappatau1schw}).
The angular velocity $\Omega_U$ given by Eq. (\ref{omUkerr}) reduces to $\Omega_U=\gamma\nu/r$. It turns out that the condition (\ref{spin_hol}) 
for a spin vector to return to its original direction after $n$ azimuthal loops is satisfied by the following values of the linear velocity
\beq
\nu_{(m,n)}^{\rm S} =\pm \left[1-\frac{n^2}{m^2}\frac{(r-3M)^2}{r(r-2M)}\right]^{1/2}\ .
\eeq
For instance 
\beq
\nu_{(m,m)}^{\rm S} =\pm \left[\frac{M(4r-9M)}{r(r-2M)}\right]^{1/2}\ ,
\eeq
implying that at $r/M=9/4$ the ZAMOs themselves are such that a spin vector rotates one full revolution during each azimuthal loop as seen from spatial infinity.

The parallel transported frame reduces to $\{\epsilon_0\equiv f_0,\epsilon_1,\epsilon_2,\epsilon_3\equiv f_3\}$, since $\sigma_B\equiv0$.
Moreover:
\begin{eqnarray}
\sigma_R&=&\gamma |{k_{\rm (lie)}}(n)_{\hat r}|\, |(\nu^2-\nu_K^2)|^{1/2}, \nonumber \\
\nu_{(m,n)}^{\rm P}&=& 
\pm \nu_K \left[\frac{r}{M}\left(1-\frac{m^2}{n^2}\right)-2 \right]^{-1/2}
\ .
\end{eqnarray}
Notice that for $m=n$ there are no orbits satisfying the holonomy condition (\ref{holonomycondition}).
The highest ratio of integers is therefore $|m/n|=1/2$, for which the values of  $\nu_{(1,2)}$  corresponding to two azimuthal loops are
\beq
\nu_{(1,2)}^{\rm P}=\pm 2M\left[(r-2M)(3r-8M)\right]^{-1/2}\ .
\eeq

\subsection{The flat spacetime limit}

The limit of vanishing mass $M=0$ of the Schwarzschild metric is the flat metric
\beq
\label{metric_flat}
\rmd  s^2 = -\rmd t^2 + \rmd r^2 
   + r^2 (\rmd \theta^2 +\sin^2 \theta \rmd \phi^2)\ .
\eeq
In this limit the ZAMOs are inertial observers ($\A_{\hat r} = 0={\theta_{\hat \phi}}(n)_{\hat r}$) and  the Lie relative curvature reduces to
\beq
{k_{(\rm lie)}}(n)_{\hat r} =-\frac1r\ .
\eeq

The Frenet-Serret frame for circular orbits in the equatorial plane $\theta=\pi/2$ is still given by Eq.~(\ref{FSframeeq}) with $\tau_2=0$ and
\beq
\label{kappatau1flat}
\kappa = -\gamma^2 \nu^2/r\ , \qquad
\tau_1 =\gamma^2 \nu/r \ . 
\eeq
The Fermi-Walker spatial frame is given by Eq.~(\ref{fwframeeq}) with Frenet-Serret triad (\ref{FSframeeq}) and first torsion given by Eq.~(\ref{kappatau1flat}).
The angular velocity $\Omega_U$ given by Eq. (\ref{omUkerr}) is still $\Omega_U=\gamma\nu/r$ and so
\beq
 |\tau_1/\Omega_U|=\gamma \ ,
\eeq 
so angle of rotation of the boosted spin after one loop with respect to the Cartesian axes (\ref{nonrotframeinfinity}), which are now globally covariant constant,
is given by the familar Thomas precession formula $2\pi (\gamma -1)$, so one has strong precession effects for sufficiently high $\gamma$.
The condition for a spin vector to return to its original direction after $n$ azimuthal loops is satisfied by the following values of the linear velocity
\beq
\nu_{(m,n)}^{\rm S} =\pm \left[1-m^2/n^2\right]^{1/2}\ .
\eeq

The parallel transported frame reduces to $\{\epsilon_0\equiv f_0,\epsilon_1,\epsilon_2,\epsilon_3\equiv f_3\}$, since $\sigma_B\equiv0$ and $\sigma_R=\gamma |\nu |/r=|\Omega_U|$.
In this case the parallel transport rotation exactly compensates the local orbital rotation, keeping the parallel transported frame constant with respect to the Cartesian axes and a parallel transported vector always returns to its original position with respect to the symmetry adapted frame after each azimuthal revolution, independent of velocity. Parallel transport is trivial in the flat Minkowski geometry, but Fermi-Walker transport is not due to the fact that commutators of boost generators in the Lorentz group Lie algebra result in rotation generators.

\section{Concluding remarks}

The relationships between the three types of physical frames associated with timelike circular orbits in stationary axisymmetric spacetimes, namely Frenet-Serret, Fermi-Walker and parallel propagated frames, have been given, and then evaluated explicitly for the Kerr and Schwarzschild spacetimes in their equatorial plane. 
The Frenet-Serret frames are locked to the symmetry adapted frames used to describe the spacetime metric.
Fermi-Walker frames encode the precession of test gyroscopes. Parallel transported frames instead describe the full parallel transport of vectors along these orbits, and have been used to explore holonomy considerations.
While these questions are more of a mathematical interest than having direct physical applications, they do reveal the rich geometry that underlies circular orbits within the general relativistic framework. 

We are happy that the pursuit of such questions brought us together with Bahram for a number of collaborative investigations that allowed us to get to know him better, as well as appreciate his formidable technical skills and special ability to harness them in analyzing physical problems.

\begin{acknowledgements}
The authors acknowledge ICRANet for support. 
\end{acknowledgements}

\end{document}